\documentclass[prb, aps,
twocolumn,
showpacs,amsmath,amssymb]{revtex4}
\usepackage{epsf}
\usepackage{graphicx}
\usepackage[english]{babel}
\usepackage{graphicx}
\usepackage{amsmath}
\usepackage{amsxtra}
\usepackage{amstext}
\usepackage{amssymb}
\usepackage{latexsym}
\usepackage{graphicx}
\usepackage{amsmath}
\usepackage{amsxtra}
\usepackage{amstext}
\usepackage{amssymb}
\usepackage{latexsym}
\usepackage[usenames]{color}

\def \be {\begin{equation}}
\def \ee {\end{equation}}
\def \BEA {\begin{eqnarray}}
\def \EEA {\end{eqnarray}}
\begin{document}

\title{ Strong magnetoresistance of disordered   graphene}
\author{P. S. Alekseev$^{1}$}
\author{A. P. Dmitriev$^{1}$}
\author{I. V. Gornyi$^{1,2}$}
\author{ V. Yu. Kachorovskii$^{1}$}

\affiliation{$^{1}$ A .F. Ioffe Physico-Technical Institute,
194021 St.~Petersburg, Russia
\\
$^{2}$ Institut f\"ur Nanotechnologie,  Karlsruhe Institute of Technology,
76021 Karlsruhe, Germany
}

\date{\today}

\begin{abstract}

We study theoretically magnetoresistance (MR) of graphene with different types of  disorder. For short-range disorder, the key parameter determining magnetotransport properties---a product of the
cyclotron frequency and scattering time---depends in graphene not only on magnetic field $H$ but also on the
electron energy $\varepsilon$. As a result, a strong, square-root in $H$,  MR arises already
within the
Drude-Boltzmann approach.
The MR is particularly pronounced near the Dirac point. Furthermore, for the same reason, ``quantum'' (separated Landau levels) and ``classical''
(overlapping Landau levels) regimes may coexist in the same sample at fixed $H.$
We calculate the conductivity tensor within the self-consistent Born approximation for the case of
relatively high temperature, when Shubnikov-de Haas oscillations are suppressed by thermal averaging.
We predict a square-root MR  both at very low and at very high $H:$ $[\varrho_{xx}(H)-\varrho_{xx}(0)]/\varrho_{xx}(0)\approx C \sqrt{H},$ where  $C$ is a temperature-dependent factor, different in the low- and strong-field limits and containing both  ``quantum''  and ``classical'' contributions.  We also find a nonmonotonic dependence of the Hall coefficient both on magnetic field and on the electron concentration.
In the case of screened  charged impurities, we  predict a strong temperature-independent  MR
near
the Dirac point. Further, we discuss the competition  between disorder- and collision-dominated mechanisms of the MR.
In particular, we find
that the square-root   MR is always established for graphene with charged impurities  in a generic gated setup at  low temperature.

\end{abstract}

\maketitle

\section{Introduction}
Study of magnetotransport in low-dimensional systems is a  powerful tool to probe the nature of disorder
and extract information about localization phenomena. In particular, measurements of magnetoresistance (MR) of two-dimensional 
(2D) electron gas in conventional semiconductors, like GaAs, allowed one to identify a variety of different regimes,
such as Drude-Boltzmann quasiclassical transport, weak localization regime and the Quantum Hall Effect
(see Refs.~\onlinecite{Ando_rev} and \onlinecite{rev_g_1} for review).

One of the simplest theoretical approaches to the  problem,
so called self-consistent  Born approximation (SCBA),  was developed in  Refs.~\onlinecite{Ando_1,Ando_2,Ando_3,Ando_4}
for 2D electrons with quadratic spectrum
 mostly for
 the case of the short-range disorder. SCBA approach ignores localization effects.
This implies that the relevant energy scale (the maximum of the chemical potential
$\mu$  and temperature $T$) is large compared to the inverse transport scattering  time $\tau_{tr},$
which coincides with the quantum scattering time $\tau_q $ for the short-range disorder. The key parameter of SCBA
is $\omega_c\tau_q,$ where  $\omega_{\mathrm{c}}$ is the cyclotron frequency.  For  weak magnetic fields,
$\omega_{\mathrm{c}} \tau_{q} \ll 1,$  calculations \cite{Ando_4} reproduce semiclassical Drude-Boltzmann result.
In the opposite limit, $\omega_{\mathrm{c}} \tau_{q} \gg 1,$ semiclassical  approach fails,
and the conductivity is given  by a sum of contributions coming from  the well separated Landau levels (LLs). \cite{Ando_1,Ando_2,Ando_3,Ando_4}

Magnetotransport in  graphene  was  studied theoretically    in Refs.~\onlinecite{Ando_1_graphene,Ando_2_graphene,theory_Sharapov,
theory_general_a,theory_general_b,anom_QHE,theory_Boltzmann,Jobst}. In Refs.~\onlinecite{Ando_1_graphene,Ando_2_graphene}
a general expression for the SCBA  conductivity tensor of graphene with short-range disorder was derived and analyzed in
detail for the case of  well separated LLs.
Other types of disorder were also discussed, \cite{theory_Sharapov,theory_general_a,theory_general_b} including
disorder  potentials having  special types of symmetries. \cite{anom_QHE}
In the collision-dominated regime (when the rate of inelastic collisions due to the electron-electron interaction exceeds the impurity-induced scattering rate),
the MR of graphene at the Dirac point was calculated in Ref.~\onlinecite{theory_Boltzmann} within the Boltzmann-equation
framework by using the relativistic hydrodynamic approach. In Ref.~\onlinecite{Jobst}, a $T$-dependent interaction-induced contribution
to the MR (on top of a substantial positive $T$-independent MR) was observed experimentally and analyzed theoretically.

A specific property of graphene   compared to conventional 2D semiconductors is the linear energy dispersion of the carriers,
\be \varepsilon_{\mathrm{k}} = \pm v \hbar k
, \label{spek}\ee resulting  in the  density of states, which increases away from the Dirac point:   \be \nu_0 (\varepsilon)=
\frac{
 N
|\varepsilon|} { 2 \pi v^2 \hbar^2}.
 \label{density}\ee
  Here $v=10^8~\text{cm/c}$ is the Fermi velocity, $\varepsilon $ is the energy counted from the Dirac point  and  $ N=2\cdot 2 =4   $ is the
spin-valley degeneracy. Corresponding wave functions are given by  $\exp(i \mathbf k\mathbf r) |\chi_{\varphi}\rangle, $
 where $|\chi_\varphi \rangle$  is the spinor  with the components   $(e^{-i \varphi/2},~ \pm e^{i \varphi/2})/\sqrt{2}$,
and $\varphi$ denotes the polar angle of the momentum $\mathbf k$.

Important consequence of the linear dispersion of graphene  is that
the cyclotron frequency  turns out to be energy-dependent: \cite{graphene_review}
\be\omega_{c}(\varepsilon)=\frac{eH}{c m(\varepsilon)}=\frac{\hbar \Omega^2}{\varepsilon}, \hspace{5mm}\text{for}~~ \varepsilon \gg \Omega .  ~~
 \label{omega-c}\ee
Here $H$ is the magnetic field and $m(\varepsilon)={\varepsilon}/{v^2}$ is the  energy-dependent cyclotron mass. The frequency 
\be
\Omega=\frac{v}{l_H}
\label{Omega}
\ee
is proportional to the distance between  the lowest LLs [see Eq.~\eqref{LL} below], where  $l_H=\sqrt{c\hbar/eH}$ is the magnetic length.
Another consequence is the energy dependence of quantum and momentum  relaxation times for the  short-range  scattering potential.
Here we by definition  assume that the  random potential  is a short-range one if  its radius $R$ satisfy the following inequalities
\be a \ll R \ll \lambda,
\label{ineqR}
\ee where $a$ is the lattice constant and $\lambda$ is the electron wavelength (this definition is different from one chosen  in Refs.~\onlinecite{Ando_potential,theory___nonrelevant_without_H}).   Such a potential does not mix valleys and can be written as\cite{Ando_potential,Ando_1_graphene}
 \begin{equation}
\label{pot} \hat{V}(\mathbf{r})= V(\mathbf r)\left(\begin{array}{cc} 1 & 0
\\
0 & 1
\end{array}\right)
,
~V(\mathbf r)=u_0 \sum_i \delta(\mathbf{r}-\mathbf{r}_{\text{i}}),\end{equation}
where summation is taken over the impurity positions.
The
correlation function of $ {V}(\mathbf{r})$  is given by $ \langle
V(\mathbf{r}) V(\mathbf{r}') \rangle= \kappa \delta(\mathbf{r}-\mathbf{r}'),$ where $\kappa=n_\text{imp} u_0^2 $ and $n_\text{imp}$ is the impurity concentration.
 Calculating by golden rule the  quantum and transport scattering times we find that they  are different and
 both  energy-dependent:  \cite{graphene_review,Ando_1_graphene}
 \begin{equation}
\label{tau0} \tau_{{q} }(\varepsilon)=\frac{\gamma
\hbar}{|\varepsilon|} \: , \;\;\;\; \tau_{{tr} }
(\varepsilon)=2\tau_{{q}} (\varepsilon) \:,
\end{equation}
where $\gamma=2 \hbar ^2 v^2/\kappa$.
 The difference between  $\tau_{{tr}}$ and  $\tau_q$ is due to weak anisotropy of the scattering arising from spinor nature of the wave functions. Indeed, the     squared scattering matrix element,
$|U_{\mathbf{k}' \mathbf{k}}|^2$ is proportional to  $|\langle\chi_{\varphi'}|
\chi_{\varphi}\rangle|^2$ and, therefore, depends on the scattering
 angle $\varphi-\varphi'.$
Below we assume that $\gamma \gg 1$ and, consequently, $\varepsilon \tau_{{q}}(\varepsilon )/
\hbar \gg 1. $ The letter inequality allows us to neglect localization effects. We also assume that disorder
does not affect density of states. This condition is  also satisfied for large $\gamma ,$  with an
exception  for exponentially small energies, \cite{theory___nonrelevant_without_H} $\varepsilon \sim
 \Delta \,e^{ - \pi \gamma /2 }$ ($\Delta$ is the
bandwidth of graphene),   which are irrelevant for this paper.  Under such assumptions the  conductivity in zero  magnetic field is given by the Drude formula:
\be
\sigma_{xx}^D  = \sigma_0 = \frac{2e^2\gamma}{\pi\hbar}, ~\text{for}~H=0.
\label{sigma0}\ee

We see that both  $\omega_{{c}}$ and $\tau_q$ are energy-dependent
and, therefore, the parameter
\be x=\omega_{{c}} \tau_q= \frac{\varepsilon_*^2}{\varepsilon^2}
 \label{x}\ee
 can be small or large at the same sample for different $\varepsilon$.
Here we  used  Eqs.~\eqref{omega-c} and \eqref{tau0}  and introduced the energy
\be \varepsilon_*=\hbar\Omega \sqrt{\gamma} \gg \hbar \Omega  ,\label{eps*}\ee
which scales as a square root of the magnetic field: \be \varepsilon_* \propto \sqrt{H} . \label{sqrt}\ee
As seen from Eq.~\eqref{x}, at sufficiently high $T$
the temperature window might include both the ``quantum''  ($x >1$) and the ``classical'' ($x  <1$) regions, so that the total conductivity might show    some peculiarities specific both for the quantum and classical transport.

In the first part of the  paper we calculate  MR  and the Hall coefficient  of graphene with the short-range disorder  assuming that  ${\rm max}(T,\mu)  \gg \hbar\Omega$
  and, consequently, the number of  filled LLs is large.

 The most interesting finding is related to the case
$\varepsilon_* \to 0 $ ($H \to 0$).
  We  demonstrate  that the dominant  contribution to the low-field resistivity comes from   the energy scale $\varepsilon_*,$ which is deep below
  $\rm{max}(T,\mu) $ in this case.
 This contribution is calculated within SCBA  based on the approach of Ref.~\onlinecite{Ando_1_graphene}. Assuming that temperature is not too small
\be
\hbar \omega_c(\varepsilon^*) \ll T
\label{ineqT}
\ee
(this condition ensures that the Shubnikov-de Haas oscillations are suppressed by energy averaging within the temperature window)
we find that  for low fields ($\varepsilon_* \ll T$) the  relative longitudinal resistivity scales as a
square root of $H$: \be \frac{\Delta \varrho_{xx}}{\varrho_{xx}(0)}\approx \frac{0.784~ \varepsilon_*}{T\cosh^2(\mu/2T)} \propto \sqrt{H}. \label{MR}\ee
Here $\Delta \varrho_{xx}= \varrho_{xx}(H)- \varrho_{xx}(0),$ and $\varrho_{xx}(0)=1/\sigma_0.$
For $T\ll \mu,$ MR is exponentially small because the energy scale $\varepsilon \sim \varepsilon^*$ is well beyond the temperature window. However, for  $T  \gtrsim \mu,$  low-field  MR is  quite large and increases with decreasing the temperature. From the side of the lowest fields, the   square root dependence \eqref{MR}  is limited by  exponentially small fields corresponding to $\varepsilon_* \sim \varepsilon_*^{{min}} \approx \Delta e^{-\pi\gamma/2} .$  Calculation of MR at $\varepsilon_*  \ll \varepsilon_*^{{min}} $ is not controlled because at such energies $\gamma$ is renormalized to unity, \cite{theory___nonrelevant_without_H} so that impurity potential becomes effectively strong and SCBA fails. One may expect that for $\varepsilon_*  \ll \varepsilon_*^{{min}} $ MR becomes parabolic which implies that one should replace factor $\varepsilon_*/T$  with $(\varepsilon_*/T)(\varepsilon_*/\varepsilon_*^{{min}})^3$ in Eq.~\eqref{MR}.

We   show that the square-root dependence of MR is also obtained   in the opposite limit of large field, $\varepsilon_* \to \infty $ ($H \to \infty$):
\be \frac{\Delta \varrho_{xx}}{\varrho_{xx}(0)}\approx \left \{ \begin{array}{l}
                                                  0.96~ \varepsilon_* T/\mu^2,~\text{for}~ T\gg\mu,~\text{and}~ \varepsilon_* \gg T^2/\mu,  \vspace{1.5mm}\\
                                                  0.68~ \varepsilon_* /\mu,~~\text{for}~ \mu \gg T,~\text{and}~\varepsilon_*\gg\mu.
                                                \end{array}
\right. \label{MR1}\ee

Further, we discuss the behavior of the Hall coefficient  $R$  as a function of the magnetic field and  the electron concentration   and demonstrate that it is  a nonmonotonic function of  both variables.
\begin{figure}[ht!]
 \leavevmode \epsfxsize=8.0cm
 \centering{\epsfbox{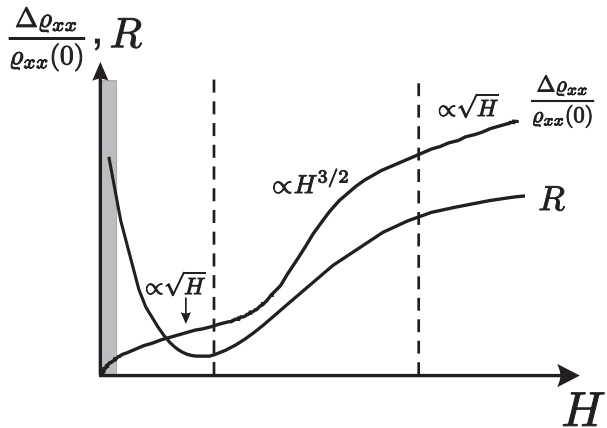}}
\caption{ Schematic plot of the dependence of the longitudinal resistivity and the Hall coefficient on the  magnetic field for $\mu \ll T$.   Region of exponentially weak fields corresponding to $\varepsilon_* < \varepsilon_{min}$ is marked by grey color. Our theory is applicable for higher fields. Vertical dashed lines correspond to $\varepsilon_*\sim T$ and $\varepsilon_* \sim T^2/\mu.$
 }
 \end{figure}

In Fig.~1 we plotted schematically the   dependence of the longitudinal resistivity and the Hall coefficient on the  magnetic field for $\mu \ll T.$ Importantly, square-root MR is predicted both for very low and very high fields.   More detailed pictures are presented below (see Figs.~4, 5, and 6).  It turns out more convenient to plot all  dependencies  not as  functions of $H$ but as function of $\varepsilon_* \propto \sqrt{H}.$ This allows us to  present the  results for $\mu \ll T$ and $\mu \gg T$ in a similar way.

 In the second part of the paper, we analyze the case of the charged impurities and   discuss the effect of the electron-electron interaction on magnetotransport. The role  of the interaction turns out to be two-fold: interaction  leads to a screening of the charged impurities and provides an additional scattering channel which limits conductivity of graphene at the Dirac point (in contrast to a conventional semiconductor).   The importance of both effects depends on the value of the  dimensionless parameter $\alpha$ which characterizes the strength of the Coulomb interaction. In this work, we  use $\alpha$ as a parameter of the theory. Specifically, in the case of charged impurities we discuss separately two cases: $\alpha \sim 1$  and $\alpha \ll 1.$  We first neglect electron-electron collisions and  demonstrate that  for weak coupling ($\alpha \ll 1$)   charged impurities yield square-root (parabolic)  MR at the Dirac point in the low- (high-) field limit.  Then we present a qualitative discussion of the role of inelastic collisions  and establish conditions of the applicability of our results in the context of the hydrodynamic treatment.  Specifically, we  compare our results with those obtained in the collision-dominated regime in Refs.~\onlinecite{theory_Boltzmann} and discuss the competition between the two mechanisms of the strong  MR at the Dirac point: (i) due to inelastic collisions\cite{theory_Boltzmann}  and (ii) due to screened charged impurities. In particular, we show that the external screening  of the charged impurities by the gate electrode favors the disorder-dominated mechanism, so that the square-root low-field  MR  is always established in a generic gated (i.e. typical for most transport experiments) setup at sufficiently low temperatures. The competition of the electron-electron collisions and scattering off the short-range impurities is also discussed and   the conditions needed   for realization of the square-root MR are presented.

\section{Basic equations}\label{sec2}
\subsection{Qualitative analysis} \label{sec2a}
 In the beginning of this section, before turning to the rigorous  calculations, it is instructive to make some  qualitative estimates clarifying the physics of the predicted square-root MR.  To this end we note
 that behavior similar to given by Eq.~\eqref{MR} may be obtained  already within the semiclassical
 Drude-Boltzmann approximation.
The main ingredient needed for obtaining the square-root MR is the
specific  energy dependence of $\omega_c$ and $\tau_{tr}$ given by Eqs.~\eqref{omega-c} and \eqref{tau0}, respectively.
Indeed, the classical  approach based on the Boltzmann kinetic equation yields
 \BEA \label{sigmaD}
 \sigma^D_{xx}(\varepsilon)&=&
 \frac{\sigma_0}{1+
 [\omega_{\mathrm{c}}(\varepsilon)\tau_{tr}(\varepsilon)]^2}
 \\&=&
 \sigma_0 \left(1- \frac{4\varepsilon_*^4}{\varepsilon^4+4\varepsilon_*^4}\right),
 \label{sigmaD1}
 \EEA
 where we used Eqs.~\eqref{tau0}, \eqref{sigma0}, and \eqref{x}.
 Importantly, the second term in the r.h.s. of Eq.~\eqref{sigmaD1} is peaked near $\varepsilon=0$  within
 the width on the order of $\varepsilon_*.$
 Let us consider for simplicity the case $\mu\ll T$ assuming that $\varepsilon_* \ll T.$
 Averaging of  Eq.~\eqref{sigmaD1} over energy within the temperature window yields two terms:
  the field-independent contribution
  $\sigma_0$ and the contribution of the peak
  \be
  \Delta\sigma_{xx}(H)\sim -(\varepsilon_*/T) \sigma_0.
  \ee
  Analogously, for the transverse conductivity we obtain
   \be
   \sigma^D_{xy}(\varepsilon)=\omega_c\tau_{tr} \sigma_{xx}^D =\sigma_0
 \frac{\varepsilon|\varepsilon|\varepsilon_*^2}{\varepsilon^4+4\varepsilon_*^4}.
 \label{sigmaDxy}
   \ee
After averaging over energy, Eq.~\eqref{sigmaDxy} yields
  \be
  \sigma^D_{xy} \sim \sigma_0\frac{\varepsilon_* \mu}{T^2} \sim \frac{\mu}{T}\left|\Delta\sigma_{xx}(H)\right| ,
  \ee
   so that the total transverse conductivity is smaller~\cite{electrons} than the field-dependent part, $\Delta\sigma_{xx}(H)$,
   of the longitudinal conductivity.  The longitudinal resistivity reads:
    \begin{eqnarray}
    \varrho_{xx}(H)&=&
    \frac{\sigma_0+\Delta\sigma_{xx}(H)}{[\sigma_0+\Delta\sigma_{xx}(H)]^2+[\sigma^D_{xy}(H)]^2}
    \nonumber \\
    &\approx& \frac{1}{\sigma_0}-\frac{\sigma_0\Delta\sigma_{xx}(H)+[\sigma^D_{xy}(H)]^2}{\sigma_0^3}.
    \label{Drude2}
    \end{eqnarray}
    Since in our case
   \be
    [\sigma_{xy}^D(H)]^2\ll \sigma_0|\Delta\sigma_{xx}(H)|,
    \label{ineq-sigmaxy}
    \ee
     we can neglect $\sigma_{xy}^D$ in Eq.~(\ref{Drude2}).
     As a result, in contrast to the conventional case of a parabolic spectrum [where $\sigma_{xy}^2=-\sigma_0\Delta\sigma_{xx}(H)$],
     the magnetic-field dependence of $\sigma_{xx}(H)$ directly translates into the MR:
     \be
     \varrho_{xx}(H)\simeq \frac{1}{\sigma_{xx}^D(H)}.
     \ee
      Hence, we find
      \be \frac{\Delta \varrho_{xx}}{\varrho_{xx}(0)} \simeq - \frac{\Delta \sigma_{xx}(H)}{\sigma_0} \sim \frac{\varepsilon_*}{T} \propto\sqrt {H} . \label{est}\ee

      It turns out, however, that this analysis yields an incorrect value of the numerical coefficient in the low-field asymptotic of MR [see discussion after Eq.~\eqref{I}]. Indeed, a purely classical Drude-Boltzmann approach is valid for $\varepsilon \gg \varepsilon_*$  and  fails at relevant energies $\varepsilon \sim \varepsilon_*$ where the LLs  start to separate.
      For   $\varepsilon \ll \varepsilon_*,$ the LLs are well separated and
the longitudinal conductivity contains the density of states squared.
After thermal averaging this also leads  to a $\sqrt{H}-$contribution to MR  which comes from the separated  LLs (see, e.g.,  Refs.~\onlinecite{Dmitriev_review} and \onlinecite{Aleiner_smooth}) and has essentially quantum nature (despite temperature is higher than inter-level distance).
 In the calculations below we use a  rigorous approach based on the SCBA, which treats both,
classical and quantum, mechanisms of the square-root-MR
on equal footing and allows one to describe crossover between the classical and quantum regions at $\varepsilon \sim \varepsilon_*$.

\subsection{Self-consistent Born Approximation for graphene with short-range disorder
}
Inequality \eqref{ineqR} ensures that disorder does not mix valleys  two equivalent valleys of graphene,  so that one may calculate the conductivity in one valley and  then simply multiply the obtained result by the factor 2. The single-valley  Hamiltonian in the perpendicular
magnetic field is given by \cite{graphene_review,Ando_potential}
\[
                 \hat{H}=\hat{H}_0  +V_{\mathrm{imp}}(\mathbf{r})\:,
\]
\[
\hat{H}_0 =v \, \hbar\left( \hat{\boldsymbol{\sigma} }
                  \cdot \left[\hat{\mathbf{k}} +\frac{e}{c \hbar}\mathbf{A}
                  (\mathbf{r}) \right]\right) \:,
\]
where  $e$ is the
absolute value of electron charge, $V_{\mathrm{imp}}(\mathbf{r})$ is
the random impurity potential.
The eigenenergies  and  eigenfunctions of $H_0$ read \be \varepsilon_n = \hbar\Omega~ {\rm sign} (n)  \sqrt{2|n|},~~
\psi_{n,k}(x,y)= {e^{-i k y}}\chi_{n} (x-kl_H^2), \label{LL}\ee
where $\Omega$ is given by Eq.~\eqref{Omega}  and
\be
\chi_{n} (x) =  \left\{
\begin{array}{l} \displaystyle
\frac{1}{\sqrt{2}}
\left[
\begin{array}{c} h_{|n|-1} (x) \\ {\rm sign}(n)~h_{|n|}(x) \end{array}
\right],~\text{for}~~n=\pm1,\pm2,\ldots
;
\vspace{2.5mm}\\
 \displaystyle \hspace{6mm}\left[
\begin{array}{c} 0 \\ h_0(x) \end{array}
\right],\hspace{20mm} \text{for}~n=0.
\end{array}\right.
\label{gr_vw}
\ee
Here $h_{n}(x)$ are the  normalized wave functions of the
harmonic oscillator with the frequency $\Omega$ and mass $\hbar/vl_H.$  As seen from Eq.~\eqref{LL}, the energy-dependent cyclotron frequency
is connected with  $\Omega$ by Eq.~\eqref{omega-c}
 while the relevant energy scale  $\varepsilon_*$ is  given by Eq.~\eqref{eps*} and
 corresponds to high LLs, so that we can  use  the  SCBA for calculation of the resistivity.

In the  SCBA, the  electron  Green function  in the short-range potential  is given by \cite{Ando_1_graphene}
\be \hat G(\varepsilon) = \frac{1}{\varepsilon - \hat H_0 -\hat \Sigma} \label{G},\ee
 where
 self-energy
is found from the following equation
 \be
 \hat \Sigma= \kappa \left \langle \mathbf r  \left| \frac{1}{\varepsilon - \hat H_0 -\hat \Sigma}  \right | \mathbf  r  \right \rangle.
 \ee
 As shown in Ref.~\onlinecite{Ando_1_graphene}, $\hat \Sigma $ is $2\times 2$ matrix having nonzero matrix elements between $\chi_n$ and $\chi_{-n}.$
However,  at high energies corresponding to high Landau levels, $\hat \Sigma $  becomes simply proportional to the unit matrix:
      $$ \hat \Sigma (\varepsilon) \approx \Sigma(\varepsilon) \left(\begin{array}{cc}        1& 0 \\
                                                                               0 & 1 \\
                                                                             \end{array}
                                                                           \right),
  ~~\text{for}~~
 \varepsilon \gg \hbar\Omega, $$
 where
 \be \Sigma (\varepsilon)=\Delta(\varepsilon) \pm i\Gamma(\varepsilon)  \label{Sig}\ee
  (signs $+$ and $-$ corresponds to advanced and retarded self-energies, respectively).
In this approximation, $\Sigma$ obeys\cite{Ando_1_graphene}
\be
\Sigma=\frac{(\hbar\Omega)^2}{\gamma}     \sum_{n=0}^{N_{max}} \frac{\varepsilon-\Sigma}{(\varepsilon-\Sigma)^2-\varepsilon_n^2}.
\label{Sig1}
\ee
Here $N_{max}$ is the ultraviolet cutoff.   The sum entering Eq.~\eqref{Sig1} can be calculated by using the identity
\be \sum_{n=0}^{N_{max}} \frac{1}{N-W}\approx\ln\left(\frac{N_{max}}{W}\right)-\pi\cot( \pi W),\label{ident}\ee
valid for $N_{max}\gg \text{Re} (W) \gg 1, ~~\text{Im}(W) \lesssim  \text{Re} (W),$ and $\text{Im}(W)>0.$  From Eqs.~\eqref{Sig1} and \eqref{ident} we  obtain
\be
\Sigma\approx\frac{\varepsilon -\Sigma}{2\gamma} \left\{   \ln \left [   \frac{2N_{max} (\hbar\Omega)^2}{(\varepsilon-\Sigma)^2}\right]-\pi   \cot\left[\pi \frac{(\varepsilon-\Sigma)^2}{2(\hbar\Omega)^2} \right] \right\}.
\label{Sig2}
\ee
 The logarithm entering Eq.~\eqref{Sig2} is a smooth function of $\varepsilon$ and leads to a linear in $\varepsilon$ correction to $\Sigma,$   which
 is irrelevant provided that $\gamma \gg \ln N_{max}$ (see also discussion of SCBA in graphene in Ref.~\onlinecite{anom_QHE}).  We   substract this correction from $\Sigma$ and for simplicity  use  the same notation $\Sigma$ for thus redefined self-energy.
 Then we find from    Eq.~\eqref{Sig2} the system of coupled equations for $\Delta$ and $ \Gamma:$
  \BEA
\label{delta}
\Delta &=& \frac{\displaystyle\Gamma_0 ~\sin
\left[{2\pi(\varepsilon-\Delta)}/{\hbar\omega_c}\right]}{\displaystyle \cosh\left[{2\pi \Gamma}/{\hbar\omega_c}\right]
-\cos\left[{2\pi(\varepsilon-\Delta)}/{\hbar\omega_c}\right]},
\\
\label{Equ_z}   \Gamma &=& \frac{\displaystyle\Gamma_0 ~\sinh\left[{2\pi \Gamma}/{\hbar\omega_c} \right]}{\displaystyle \cosh\left[{2\pi \Gamma}/{\hbar\omega_c}\right]
-\cos\left[{2\pi(\varepsilon-\Delta)}/{\hbar\omega_c}\right]},\EEA
where $\Gamma_0(\varepsilon)=\hbar/2\tau_q(\varepsilon).$
We also  normalize $\Gamma$ by its value at zero magnetic field  introducing the quantity
 \be
 z(\varepsilon) = \frac {\Gamma(\varepsilon)}{\Gamma_0(\varepsilon)}=\frac{\nu(\varepsilon)}{\nu_0(\varepsilon)}, \label{z}
 \ee
   Here $\nu(\varepsilon)$ is the density of states in the magnetic field.
 The solution of  Eqs.~\eqref{delta} and \eqref{Equ_z} can be found analytically  in the limiting cases $\varepsilon \gg \varepsilon_* ~(x\ll 1)$ and $\varepsilon \ll \varepsilon_* ~(x\gg1):$
 \be \label{delta1}\Delta=\left\{
\begin{array}{l} \displaystyle 2\Gamma_0 e^{-\pi/x} \sin\left( \frac{2\pi\varepsilon}{\hbar\omega_c}\right)
, ~\text{for}~~ \varepsilon \gg \varepsilon_* ;
\vspace{1.5mm}\\
\displaystyle \sum_n \vartheta\left(\frac{\varepsilon-\varepsilon_n}{\Gamma_n}\right)\frac{\varepsilon}{2}
, ~\text{for}~~
\varepsilon \ll \varepsilon_* ,
\end{array}\right.\ee
 \be \label{z_in_High}
z
\approx\hspace{-1mm}\left\{
\begin{array}{l}
\displaystyle
\hspace{-2mm} 1+ 2a e^{-\pi/x} + 2(2a^2-1)\left(\hspace{-1mm}1-\frac{2 \pi  }{x}
\hspace{-1mm}\right)e^{-2\pi/x},
~\text{for}~ \varepsilon \gg \varepsilon_*;
\vspace{1.5mm}\\
\displaystyle
\hspace{-1mm} \sqrt{\frac{2x}{\pi}}\:
\sum_n\vartheta\left(\frac{\varepsilon-\varepsilon_n}{\Gamma_n}\right)\sqrt{1-\left(\frac{\varepsilon-\varepsilon_n}{\Gamma_n}\right)^2},
~\text{for}~~
\varepsilon \ll \varepsilon_*.
\end{array}\right.
\ee
Here
\be
a=\cos\left(  \frac {2\pi \varepsilon}{\hbar\omega_c}\right),
\label{a}
\ee
$\vartheta(y)$ is  equal to unity (zero)     for $|y|<1$ ($|y|>1$) and
\be
\Gamma_n=\Gamma(\varepsilon_n)= \hbar\sqrt{
\frac{2\omega_c (\varepsilon_n)}{\pi\tau_q(\varepsilon_n)}}
.
\label{Gamma-n}
\ee
In the first line of Eq.~\eqref{z_in_High} we expanded $z$ up to the second order with respect to $\exp(-\pi/x).$
As seen from Eqs.~\eqref{omega-c}, \eqref{tau0}, and \eqref{Gamma-n}, $\Gamma_n$ actually does not depend on $n$ for the case of short-range disorder which we concern with:
$ \Gamma_n=\Gamma_\Omega=const.$
 Here \be\Gamma_\Omega=\hbar\Omega\sqrt{\frac{2}{\pi \gamma}}. \ee Hence,  the widths of different LLs are
equal, but the distance $\hbar \omega_c
$ between neighbor LLs decreases with increasing
$\varepsilon $. We also  see that self-energy changes periodically with $\varepsilon$ on the scale $\hbar \omega_c.$  Both the period and the shape of these oscillations      slowly depend on the energy due to energy dependence of $x.$
The density of electron states, $\nu(\varepsilon),$ following from Eqs.~\eqref{z} and \eqref{z_in_High} is
plotted in Fig.~2 for $\varepsilon_* \ll     \mu .$   At  low energies,  $x \ll 1,$  LLs are well separated, while for $x \gg 1$ density of states  is given by zero-field density, Eq.~\eqref{density}, up to exponentially small oscillating terms.

\begin{figure}[ht!]
 \leavevmode \epsfxsize=8.0cm
 \centering{\epsfbox{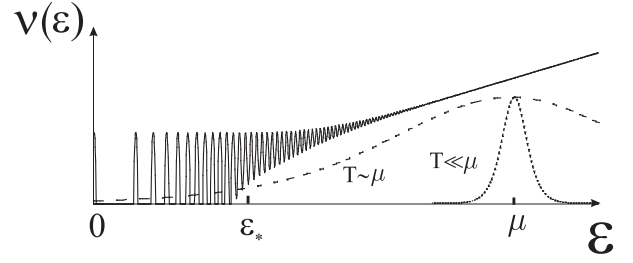}}
\caption{ Density of the electron states in graphene with short range
disorder (solid line) and the derivative of the Fermi function, $-\partial n_F/\partial\varepsilon$ , for
the temperatures $ T \ll \mu$ and $ T \sim \mu$ (dashed lines).
 }
 \end{figure}

\section{Calculation of the conductivity}
 The conductivity tensor  is given by thermal averaging of the energy-dependent tensor $\sigma_{ij}(\varepsilon)$   \begin{equation}
\label{sigma} \sigma_{ij} = \int \limits _{-\infty}  ^ {\infty }  d
\varepsilon \left[ - \frac{\partial n_F (\varepsilon) }{\partial
\varepsilon}
 \right ] \sigma_{ij}(\varepsilon),
\end{equation}
where $n_F(\varepsilon)$ is the Fermi distribution function.

 The longitudinal conductivity $\sigma_{xx}(\varepsilon)$ is calculated   by summation of the ladder diagrams.\cite{Ando_1_graphene}  The result is given by Eq.~(4.13) of Ref. \onlinecite{Ando_1_graphene}. Using Eq.~\eqref{ident} one may  rewrite this result in terms of $z(\varepsilon)$
 \begin{equation}
\label{sigma_g} \sigma_{xx}(\varepsilon)=\sigma_0
 \frac{z(\varepsilon)^2}{z(\varepsilon)^2+
 [\omega_{\mathrm{c}}(\varepsilon)\tau_{tr}(\varepsilon)]^2}\:.
\end{equation}

 For high energies $x \to 0,$  so that $z\to 1$ [see Eq.~\eqref{z_in_High}] and we obtain Drude result, Eq.~\eqref{sigmaD}.
    For $x \gg 1,$ we find from
  Eqs. (\ref{sigma_g}) and (\ref{z_in_High}) the conductivity near $n$-th LL \cite{Ando_1_graphene}
\begin{equation}
\label{And}
 \sigma_{xx}(\varepsilon)
 =\left\{
\begin{array}{l} \displaystyle
 \frac{2e^2 n}{\pi^2 \hbar}
\left[1-\left(\frac{\varepsilon-\varepsilon_n}{\Gamma_\Omega}\right)^2\right]
\:,\;\;\;\text{for}~ |\varepsilon-\varepsilon_n|<\Gamma_\Omega,
\vspace{1.5mm}\\
\displaystyle 0\:,\;\;\;\text{for}~|\varepsilon-\varepsilon_n|>\Gamma_\Omega.
\end{array}\right.
\end{equation}

   We note that Eq.~\eqref{sigma_g} may be obtained from Eq.~\eqref{sigmaD} by replacement of $1/\tau_{tr}(\varepsilon)$ with  $z(\varepsilon)/\tau_{tr}(\varepsilon).$  Hence,  the only difference of  the SCBA result
   compared to  the Drude one is       the renormalization of the density of states given  by  Eq.~\eqref{z}.

The calculation of $\sigma_{xy}(\varepsilon)$  is more subtle. Simplest approximation  based on summation of the ladder diagrams leads to Drude-like formula with renormalized  scattering rate,
\be\label{sigmaxy0}
\sigma_{xy}^{\uppercase\expandafter {\romannumeral 1}} (\varepsilon)=\sigma_0
\frac{\omega_c(\varepsilon)\tau_{tr}(\varepsilon)z(\varepsilon)}
{z^2(\varepsilon)+[\omega_c(\varepsilon)\tau_{tr}(\varepsilon)]^2},
\ee
in a full analogy with Eq.~\eqref{sigma_g}.  In  fact, there also exists another contribution to the transverse conductivity, which is expressed via thermodynamical properties of the electron gas:\cite{ando-xy, streda}
\be \sigma_{xy}^{\uppercase\expandafter {\romannumeral 2}}(\varepsilon)=ec[\partial n/\partial H]_\varepsilon.
\label{sigmaxy2}
\ee
Here $n=n(\varepsilon,H) $ is the electron concentration    in magnetic field for  zero temperature and chemical potential coinciding with $\varepsilon.$ The derivative over $H$ is taken for fixed $\varepsilon.$ Hence,
\be
\sigma_{xy} (\varepsilon)=\sigma_{xy}^{\uppercase\expandafter {\romannumeral 1}} (\varepsilon) +\sigma_{xy}^{\uppercase\expandafter {\romannumeral 2}} (\varepsilon).
\ee
In fact, $\sigma_{xy}^{\uppercase\expandafter {\romannumeral 2}} (\varepsilon)$ yields essential contribution to $\sigma_{xy} (\varepsilon)$ for $\varepsilon \ll \varepsilon_*$ when LLs are well separated (in the opposite case of overlapping LLs $\sigma_{xy}^{\uppercase\expandafter {\romannumeral 2}} (\varepsilon)$ is exponentially small).

Let us now do the integral in  Eq.~\eqref{sigma}. As follows from Eq.~\eqref{z_in_High}, the dependence $\sigma_{ij}(\varepsilon)$   contains fast oscillations on the scale $\hbar \omega_c,$   the shape and the period of the oscillations being   energy-dependent due to energy dependence of $x.$
 Therefore, for relatively high temperature  the integration in Eq.~\eqref{sigma} may be performed in two steps.
First,
we average $\sigma_{ij}(\varepsilon)$ in Eq. (\ref{sigma}) by the
energy interval $\delta  \varepsilon$, such that $
\varepsilon \gg \delta  \varepsilon \gg \hbar
\omega_{\mathrm{c}} (\varepsilon) .$
 Such an averaging ``filters'' the Shubnikov-de Haas oscillations and results in a
smooth function $ \bar{\sigma}_{ij}(\varepsilon) $.  One can also show that after averaging
$\bar \sigma_{xy}^{\uppercase\expandafter {\romannumeral 2}}$  can be neglected \cite{koechto_0} (provided that $\gamma \gg 1$) 
 both for overlapping \cite{koechto} ($\varepsilon \gg \varepsilon_* $) and for separated ($\varepsilon \ll \varepsilon_* $)  LLs.
Hence below we use
$$\bar \sigma_{xy}(\varepsilon)\approx \bar \sigma_{xy}^{\uppercase\expandafter {\romannumeral 1}}(\varepsilon).$$

It is convenient to write  $\bar{\sigma}_{ij}$ in the following form
\begin{equation}\label{def}
\bar{\sigma}_{ij}(\varepsilon)  = \sigma_{ij}^
{\mathrm{D}}(\varepsilon) \: \eta_{ij}(\varepsilon)\:,
\end{equation}
where $\eta_{ij}(\varepsilon)$ are  dimensionless
factors. From Eqs.~\eqref{z_in_High}, \eqref{a},  \eqref{sigma_g}, and \eqref{sigmaxy0}  we  find asymptotical behavior  of $\eta_{ij}:$
\begin{equation}
\label{appr-eta-xx}
 \eta_{xx}
\approx \left\{
\begin{array}{l}
\displaystyle  \hspace{-2mm}  1 - 24x^2 e^{- 2 \pi /x  }=1-\frac{24\varepsilon_*^4}{\varepsilon^4}e^{{-2\pi\varepsilon^2}/{\varepsilon_*^2}},  ~\text{for}~ \varepsilon \gg \varepsilon_*,
\\
 \displaystyle \hspace{-2mm} C_1 \sqrt{x}=C_1{\varepsilon_*}/{\varepsilon}, ~ \text{for}~ \varepsilon \ll \varepsilon_*,
\end{array} \right.
\end{equation}
\begin{equation}
\label{appr-eta-xy}
 \eta_{xy}
\approx \left\{
\begin{array}{l}
\displaystyle 1 + 2 e^{- 2 \pi /x  }=1+2e^{{-2\pi\varepsilon^2}/{\varepsilon_*^2}} \, \:, ~\text{for}~ \varepsilon \gg \varepsilon_*,
\\
 \displaystyle 1 \, \:,~ \text{for}~ \varepsilon \ll \varepsilon_*.
\end{array} \right.
\end{equation}
Here $C_1 = 8\sqrt {2} / 3 \pi \sqrt{\pi} \approx 0.68$. For arbitrary
values of $\varepsilon$ the factors $\eta_{ij}$ have been calculated numerically.
The dependencies  $\bar{\sigma}_{xx}(\varepsilon)$ and $\bar{\sigma}_{xy} (\varepsilon)$  are plotted schematically
in Fig.~3.  In this picture we took into account that  $\sigma_{xx}(\varepsilon)=\sigma_{xx}(-\varepsilon),\sigma_{xy}(\varepsilon)=-\sigma_{xy}(-\varepsilon)$ because of the particle-hole symmetry.
In the high-energy asymptotics we keep exponentially small terms [proportional to $\exp(-2\pi/x)$], since they are important in the calculation of MR in the limit of low temperature and low field (see below). It is worth also noting that averaged longitudinal conductivity is not exactly zero at $\varepsilon=0$ but saturates  at $\varepsilon \sim \hbar \Omega$ at a quite    small value:  $ \bar \sigma_{xx}  \sim \sigma_0/\sqrt{\gamma} \ll \sigma_0 .$
 \begin{figure}[ht!]
 \leavevmode \epsfxsize=8.0cm
 \centering{\epsfbox{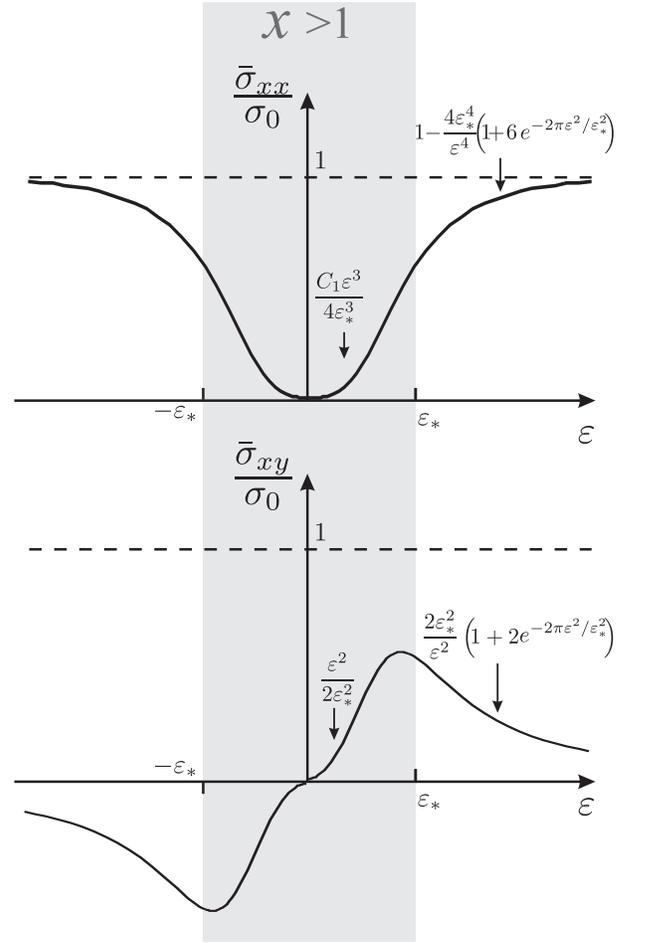}}
 \caption{
 Dependence of $\bar{\sigma}_{xx}$ and $\bar{\sigma}_{xy}$ on energy. The region $x>1$ is marked by grey color.
 }
 \end{figure}

As a second step, we  calculate $\sigma_{ij}$ by replacing $\sigma_{ij}(\varepsilon)$ with $\bar{\sigma}_{ij}(\varepsilon)$ in Eq.~\eqref{sigma}:
\begin{equation}\label{main_result}
\sigma_{ij}=
 \int d\varepsilon\: \left[ - \frac{
\partial
 n_F\left(\varepsilon\right)}{\partial
 \varepsilon
}\right]\:\bar{\sigma}_{ij}(\varepsilon),
\end{equation}
and substitute thus obtained $\sigma_{ij}$  into expression for the longitudinal resistivity
\be
\varrho_{xx}=\frac{\sigma_{xx}}{ \sigma_{xx}^2 +\sigma_{xy}^2}.
\label{rhoxx}\ee

The result of calculations depends on relation between  tree relevant energies $T,\mu$ and $\varepsilon_*.$
The magnetic
field dependence of  $\varrho_{xx}$ is encoded in the square-root scaling of  $\varepsilon_*.$ Below we discuss separately the cases of low and high temperatures.

\subsection{Low temperatures, $ T \ll \mu.$}
We start with discussing of the high-field limit, $\varepsilon_* \gg \mu.$ Since integral over energy in Eq.~\eqref{main_result} is concentrated in the narrow temperature window near $\varepsilon=\mu$ we can use low energy-asymptotics for $\bar{\sigma}_{xx}$ and $\bar{\sigma}_{xy},$ see Fig.~3. Doing so, and replacing in Eq.~\eqref{main_result} $\left[ - {
\partial
 n_F\left(\varepsilon\right)}/{\partial
 \varepsilon
}\right]$  with $\delta(\varepsilon-\mu),$  we find \be \sigma_{xx}\approx\sigma_0 \frac{C_1\mu^3}{4\varepsilon_*^3},\hspace{2mm}   \sigma_{xy}\approx\sigma_0 \frac{\mu^2}{2\varepsilon_*^2}.  \ee
We see that $\sigma_{xx}\ll \sigma_{xy}.$ Therefore, $\varrho_{xx}\approx\sigma_{xx}/ \sigma_{xy}^2$ and
\be
\frac{\Delta \varrho_{xx}}{\varrho_{xx}(0)} \approx \frac{C_1\varepsilon_*}{\mu} \propto \sqrt{H} ,\hspace{2mm} \text{for} \hspace{2mm} \varepsilon_* \gg \mu.
\label{rhoxx1}\ee

Next we consider the opposite case $\varepsilon_* \ll \mu.$ In this case, there are different competing contributions to the MR.  First contribution is obtained quite analogously to high-field limit by  replacing derivative from Fermi distribution with delta function.    In this approximation, $\sigma_{xx}=\bar\sigma_{xx}(\mu),\sigma_{xy}=\bar\sigma_{xy}(\mu).$  As follows from Eqs.~\eqref{appr-eta-xx} and \eqref{appr-eta-xy}, in the limit of large $\mu$ these conductivities differ from Drude values by  exponentially small terms only.  It is well known that in the Drude-Boltzmann approximation MR is absent in the limit of low $T.$   Hence, MR should be exponentially small. Indeed, substituting $\bar\sigma_{xx}(\mu)$ and $\bar\sigma_{xy}(\mu)$ into Eq.~\eqref{rhoxx}, using Eqs.~\eqref{def}, \eqref{appr-eta-xx}, and \eqref{appr-eta-xy},  and keeping terms  on the order of $\exp(-2\pi/x)$ we obtain   \be
\frac{\Delta \varrho_{xx}}{\varrho_{xx}(0)} \approx    \frac{8 \varepsilon_*^4}{\mu^4} e^{-2\pi \mu^2/\varepsilon_*^2},\hspace{2mm} \text{for} \hspace{2mm} \varepsilon_* \ll \mu.
\label{rhoxx-exp}\ee
There are two other contributions which may compete with this exponentially small result. Both contributions arise due to the finite value of  temperature, which was in fact assumed to be zero while deriving Eq.~\eqref{rhoxx-exp}. Let us now take into account that the  function
\be  - \frac{\partial n_F (\varepsilon) }{\partial
\varepsilon}=\frac{1}{4T\cosh^2[(\varepsilon-\mu)/2T]} \ee
is peaked near $\varepsilon=\mu$ within a finite width on the order of $T.$
First of all,       there  exists a  correction to the MR due
to a small variation of  $x(\varepsilon)$ within the temperature window.  To find this correction we put $\eta_{ij}\approx 1,$  expand $\bar \sigma_{ij} (\varepsilon)$ near $\varepsilon=\mu$ up to the second order with respect to $\varepsilon-\mu,$ calculate integral Eq.~\eqref{main_result} and use Eq.~\eqref{rhoxx}.  As a result, we get     a  quadratic-in-$H$   MR
\begin{equation}\label{rho-quadratic}
\frac{ \Delta\varrho_{xx}}{\varrho_{xx}(0)} = \frac{C_2 T^2 \varepsilon_*^4}{\mu^6} \propto H^2,
\end{equation}
where $C_2\approx 16\pi^2/3.$
This correction becomes lager than Eq.~ \eqref{rhoxx-exp} for relatively weak fields such that $\varepsilon_* \ll \varepsilon_1,$
where
\be \varepsilon_1=\frac{\sqrt{2\pi} \mu}{\ln^{1/2}(8\mu^2/C_2 T^2)}.\ee
At very  low  magnetic fields, another contribution comes into play, namely, the contribution to integral Eq.~\eqref{main_result} from the  energies $\varepsilon \sim \varepsilon^* $ which are well beyond the temperature window.  To find this contribution we first  write $\bar{\sigma}_{xx}(\varepsilon)$ in the following way
\be
\bar{\sigma}_{xx}(\varepsilon)  =
 \sigma_0[1-f(\varepsilon/\varepsilon_*)].
\ee
 The function $f(y)$ has  maximum at $y=0$ and decays when $y$ becomes larger than $1,$ or, equivalently, $\varepsilon $  becomes larger than $\varepsilon_*$ (see  Fig.~3)
  \begin{equation}
\label{f}
 f(y)\approx
 \left\{
\begin{array}{l} \displaystyle   \frac{4}{y^4},
     \hspace{4mm} \text{for}~ y \gg 1~  (\varepsilon \gg \varepsilon_*);
\vspace{1.5mm}\\
\displaystyle 1-C_1y^3/4 , ~ \text{for}~ y\ll 1~ (\varepsilon\ll \varepsilon_*).
\end{array}\right.
\end{equation}
Here we neglected exponentially small corrections to $f$ at large $\varepsilon$ and introduced dimensionless variable $y=\varepsilon/\varepsilon_*=1/\sqrt{x}$. Let us now assume that  $\varepsilon^*$ is smaller than $T .$ Then one can replace  $\left[ - {\partial n_F (\varepsilon ) }/{\partial
\varepsilon} \right ]$ with its value at zero energy $-  n_F' (0 ) ,$ while calculating the contribution to $\sigma_{xx}$ coming from the region  $\varepsilon \sim \varepsilon^* $. Doing so,
we obtain
\be
\sigma_{xx}=\sigma_0\left[1-  2C_3\varepsilon_* |n_F '(0 )|   \right],
\label{sqh}
\ee
where  dimensionless  constant $C_3$ is given by  the integral
\be
C_3= \int \limits_0^\infty dy~{ f(y)}=1.568.
\label{I}
\ee
It is instructive to compare the obtained result with the  ``purely classical'' calculation  which ignores existence of LLs.  On the formal level,
such an approximation implies replacement $\eta_{ij}$ with unity in Eq.~\eqref{main_result}.  Simple calculation (see also discussion in Sec.~\ref{sec2a})  shows that thus calculated conductivity can be  presented in the same form  as Eq.~\eqref{sqh}, where one should replace $C_3$ with
\be C_3^D= \int \limits_0^\infty dy \left( 1-\frac{1}{1+4x^2}\right)= \int \limits_0^\infty dy  \frac{4}{y^4+4}= \pi/2=1.571.\ee
This value is close but different from $C_3,$ so that discreteness of LLs should be   taken into account.

Now we are ready to calculate resistivity correction.  In the regime under discussion ($\varepsilon_* \ll T \ll \mu$),    $\sigma_{xx} \gg \sigma_{xy},$ so that  $\varrho_{xx}\approx 1/\sigma_{xx}$  and
\be
\label{rhosmall}
\frac{ \Delta\varrho_{xx}}{\varrho_{xx}(0)} \approx 2C_3\varepsilon_* |n_F '(0 )|  \propto \sqrt{H}.
\ee
Since  $T \ll \mu,$ we conclude that MR is exponentially small:
\be
\label{rhosmall1}
\frac{ \Delta\varrho_{xx}}{\varrho_{xx}(0)} \approx \frac{2C_3\varepsilon_*}{T} e^{-\mu/T}.
\ee
Comparing  Eq.~\eqref{rhosmall1} with Eq.~\eqref{rho-quadratic}, we see that the former contribution dominates when
$\varepsilon_* \ll \varepsilon_2 ,$ where
\be \varepsilon_2 \sim\frac{\mu^2}{T} e^{-\mu/3T}\ll T.\ee
Therefore,  in accordance with our assumption, $\varepsilon_* \ll T$ in the regime when Eq.~\eqref{rhosmall1} dominates.

Looking now more attentively at the above derivation one concludes that Eq.~\eqref{rhosmall} is valid at arbitrary relation between $T$ and $\mu$ provided that $\varepsilon_* \to 0$ (but $\varepsilon_* > \varepsilon_*^{min} $).  Hence,   the low-field MR is given by
 \be
\frac{ \Delta\varrho_{xx}}{\varrho_{xx}(0)} \approx \frac{C_3\varepsilon_*}{2T\cosh^2(\mu/2T)}   \propto \sqrt{H}.
\label{low-field1}
\ee

\subsection{High temperatures, $ \mu \ll T.$}
In this case, the low field asymptotics of MR is realized at $\varepsilon_*\ll T$ and  is given by Eq.~\eqref{low-field1}, where one can put $\mu=0$
\be
\frac{ \Delta\varrho_{xx}}{\varrho_{xx}(0)} \approx \frac{C_3\varepsilon_*}{2T} \propto \sqrt{H}/T, ~\text{for} ~   \varepsilon_* \ll T.
\label{low-field}
\ee
Hence, near the Dirac point resistivity correction   scales as $\sqrt{H}$ at $H\to 0,$   and increases (for fixed $H$) with decreasing the temperature.

Let us now consider larger fields,  $\varepsilon_*  \gg T. $ In this case, one can use low-energy asymptotics, \be \bar \sigma_{xx}(\varepsilon)\approx  \frac{C_1}{4}~\frac{ |\varepsilon|^3}{\varepsilon_*^3}, ~~ \bar \sigma_{xy}(\varepsilon)
\approx \frac{\sigma_0}{2} ~\frac{\varepsilon |\varepsilon|}{\varepsilon_*^2}. \ee
Thermal averaging yields
\BEA
\nonumber &&\sigma_{xx}=
\\
\nonumber
&& \frac{\sigma_0 C_1}{16}\int_0^\infty d\varepsilon  \frac {\varepsilon^3}{\varepsilon_*^3T}  \left[\frac{1}{\cosh^2 \left(\frac{\varepsilon-\mu}{2T }\right)}+ \frac{1}{\cosh^2 \left(\frac{\varepsilon+\mu}{2T }\right)}\right]
\\
&&\approx  \frac{9\sigma_0C_1 \zeta(3)}{4} \frac{T^3}{\varepsilon_*^3},
\label{sigmaxx1}
\\&&\sigma_{xy}= \nonumber
\\\nonumber
& & \frac{\sigma_0 }{8}\int_0^\infty d\varepsilon  \frac {\varepsilon^2}{\varepsilon_*^2T}  \left[\frac{1}{\cosh^2 \left(\frac{\varepsilon-\mu}{2T }\right)}- \frac{1}{\cosh^2 \left(\frac{\varepsilon+\mu}{2T }\right)}\right]
\\&&\approx
2\sigma_0\ln2 ~\frac{\mu T}{\varepsilon_*^2},
\label{sigmaxy1}
\EEA
where $\zeta$ is Riemann zeta function, $\zeta(3)\approx 1.2.$ From Eqs.~\eqref{sigmaxx1} and \eqref{sigmaxy1} we find
\BEA
\frac{ \Delta\varrho_{xx}}{\varrho_{xx}(0)} &&\approx \frac{4}{9C_1\zeta(3)}~\frac{\varepsilon_*^3}{T^3}~\frac{1}{\displaystyle 1+\left[\frac{8\ln2}{9C_1\zeta(3)}\right]^2\frac{\mu^2\varepsilon_*^2}{T^4}}
\nonumber
\\&& \approx \left\{\begin{array}{l}
                      C_4 \varepsilon_*^3/T^3,~ \text{for}~ T\ll\varepsilon_* \ll T^2/\mu \vspace{2mm}; \\
                      C_5\varepsilon_* T/\mu^2,~\text{for}~   T^2/\mu \ll \varepsilon_*.
                    \end{array}\right.
\label{high-field1}
\EEA
  Here $C_4=4/9C_1\zeta(3)\approx 0.54,$ $C_5=9 C_1 \zeta(3)/16 (\ln 2)^2\approx 0.96.$

  The results of calculations are summarized in the Fig.~4.

\begin{figure}[ht!]
 \leavevmode \epsfxsize=8.0cm
 \centering{\epsfbox{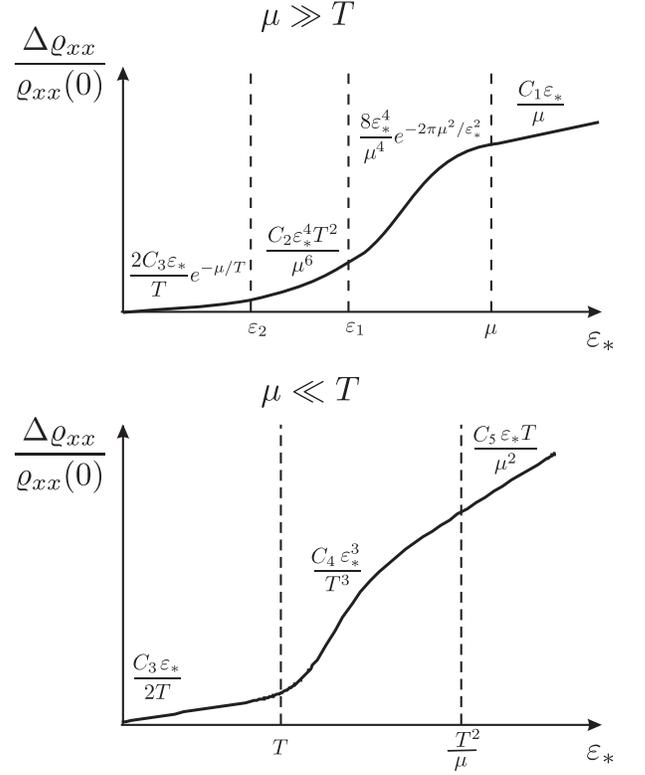}}
 \caption{
Resistivity of graphene for the cases $  T \ll \mu$  and  $T \gg \mu$ as a function  of $\varepsilon_* \propto \sqrt{H}$.
 }
 \end{figure}

\section{Hall coefficient}
Using the equations derived above one can   easily calculate the transverse  resistivity $\varrho_{xy}= \sigma_{xy}/(\sigma_{xx}^2+\sigma_{xy}^2)$ and the Hall coefficient
\be
R=\frac{\varrho_{xy}}{H}.
\ee
Below we discuss the dependence  of $R$ on $H$ and on the electron concentration at zero field,
\be
n=\int_{-\infty}^{\infty} n_F(\varepsilon)\nu_0(\varepsilon)d\varepsilon -\int_{-\infty}^{0}  \nu_0(\varepsilon)d\varepsilon.
\label{n}
\ee
 The second term in the r.h.s. of Eq.~\eqref{n}
 is  the concentration  of background electrons  which compensate the positive charge of the donors in the   neutrality point. One can easily check that $n=0$ for $\mu=0.$

 For  $T \ll \mu,$ simple calculation yields that   the Hall coefficient  up to  small corrections  is given by the conventional expression
\be
R=\frac{1}{ecn}, \label{R}
  \ee
both at very low ($\varepsilon_* \ll \mu$)  and at very high ($\varepsilon_* \gg \mu$) magnetic field.    For $\varepsilon_* \sim \mu,$ Eq.~\eqref{R} is valid up to a numerical factor on the order of unity ($R\sim 1/ecn$).     Concentration entering Eq.~\eqref{R} is given by \be n \approx
\int_0^\mu \nu_0(\varepsilon')d\varepsilon'=\frac{
 N
\mu |\mu| } { 4 \pi v^2 \hbar^2}. \ee

Consider now vicinity of the Dirac point,  $\mu \ll T.$
  In this case,
  \be
n \approx \frac{N \mu T\ln2  }{\pi  v^2 \hbar^2},
\label{nDirac}\ee
 while the transverse conductivity   is given by
\BEA
 \nonumber
\sigma_{xy}&&=\int_0^\infty d\varepsilon   \frac{\bar \sigma_{xy}(\varepsilon)}{4T} \left[\frac{1}{\cosh^2(\frac{\varepsilon-\mu}{2T})}
-\frac{1}{\cosh^2(\frac{\varepsilon+\mu}{2T})}  \right]
\\
&&\approx \frac{\mu}{2T^2} \int_{0}^{\infty}d\varepsilon \bar \sigma_{xy}(\varepsilon)\frac{\sinh(\frac{\varepsilon}{2T})}{\cosh^3(\frac{\varepsilon}{2T})}.
\label{sigmaxy3}
\EEA

\begin{figure}[ht!]
 \leavevmode \epsfxsize=8.0cm
 \centering{\epsfbox{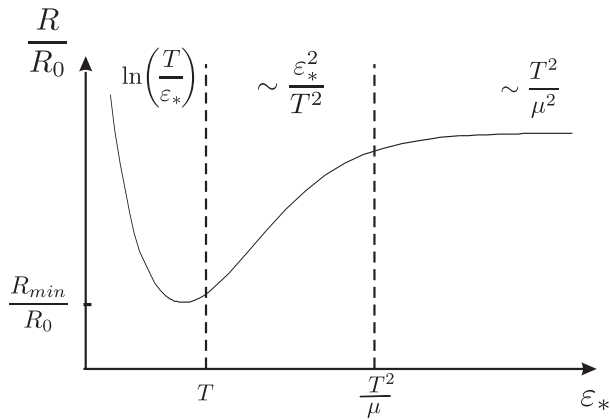}}
 \caption{
Dependence of the Hall coefficient, $\varrho_{xy}/H,$ on $\varepsilon_* \propto \sqrt{H}$  at $T \gg \mu.$
 }
 \end{figure}

For $\varepsilon_* \ll T,$   the main contribution to $\sigma_{xy}$ comes from the energy interval $\varepsilon_* <\varepsilon< T,$ where integral in Eq.~\eqref{sigmaxy3} is logarithmically divergent.   Therefore,  one may  use large-$\varepsilon$ asymptotic, $\bar \sigma_{xy} \approx 2\sigma_0\varepsilon_*^2/\varepsilon^2,$    and  calculate the integral  in the limits $\varepsilon_*$ and $T.$ Doing so we find with the  logarithmic precision: $\sigma_{xy} \approx \sigma_0 (\mu\varepsilon_*^2/2T^3) \ln(T/\varepsilon_*).$ The longitudinal conductivity  is given by  Eq.~\eqref{sqh} where one can neglect small correction proportional to $\varepsilon_*/T,$ thus writing $\sigma_{xx} \approx \sigma_0.$
Using these equations we find
\be
R= R_0~\ln\left(\frac{T}{\varepsilon_*} \right),
\label{R01}\ee
where
\be
R_0=\frac{\pi\mu v^2\hbar^2}{e c NT^3} \sim\frac{n}{ecn_T^2},
\label{R1}\ee
where $n_T \sim N T^2/v^2\hbar^2$ is the electron concentration for $\mu \sim T.$
Hence, the Hall coefficient logarithmically increases with decreasing the magnetic field.   Above we noticed that our calculations are valid up to the exponentially small fields where $\varepsilon_* \sim \Delta e^{-\pi \gamma /2}.$ Therefore,  the maximal value of the  $\ln(T/\varepsilon_*)$
is limited by   $\pi \gamma /2 -\ln(\Delta/T).$

In the opposite case, $\varepsilon_* \gg T,$ the conductivity tensor is given by  Eqs.~\eqref{sigmaxx1} and \eqref{sigmaxy1} which yield for the  Hall coefficient the following expression
\be
R= R_0 \frac{ C_6 {\varepsilon_*^2}/{T^2}}{\displaystyle 1+C_7{\mu^2\varepsilon_*^2}/{T^4}},
\label{R2}
\ee
where $C_6={64 \ln2}/{[9C_1\zeta(3)]^2} \approx 0.82$ and  $C_7=[{8\ln2}/{9C_1\zeta(3)}]^2 \approx 0.57.$
%
We see that  the Hall coefficient linearly increases with the magnetic field, $R \propto \varepsilon_*^2 \propto H ,$ in the interval $T \ll \varepsilon_* \ll T^2/\mu$  and saturates when $ \varepsilon_* $ becomes  larger than $ T^2/\mu.$       From Eqs.~\eqref{R1} and \eqref{R2} we conclude that $R$ is non-monotonic function of the magnetic field and has a minimum (for positive $\mu$) at magnetic fields corresponding to $\varepsilon_* \sim T.$ The minimal value of $R$ is given by $R_{min} \sim R_0$. The dependence of $R$ on $\varepsilon_*$ is plotted schematically in Fig.~5.

Measurements of the Hall coefficient  are usually used for extracting the density of the carriers with the help of conventional expression \eqref{R}. From  Eqs.~\eqref{nDirac}, \eqref{R01},  and \eqref{R2} we see that such a procedure fails near the Dirac point.
Let us, therefore, discuss the dependence of $R$ on $n$ in detail (see also Ref.~\onlinecite{drag}). The analysis of the above equations shows that   $R$ is a nonmonotonic function of $n$  both at low and high fields: it turns to zero for $n=0,$  has a maximum at certain $n=n_m $ and decays as $1/n$ according to Eq.~\eqref{R} for $n\to \infty $ (see Fig.~6).
\begin{figure}[ht!]
 \leavevmode \epsfxsize=8.0cm
 \centering{\epsfbox{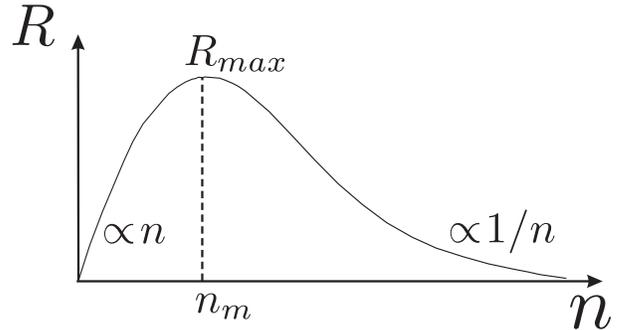}}
 \caption{
Dependence of the Hall coefficient on electron concentration
 }
 \end{figure}

As follows from Eqs.~\eqref{nDirac}, \eqref{R01}, and  \eqref{R1}, at low fields or, equivalently,  high temperatures ($\varepsilon_* \ll  T$), the Hall coefficient linearly increases with $n$ at  $n\ll n_T$ ($\mu \ll T$),
\be R \sim  \frac{ n }{ec n^2_T}\ln(T/\varepsilon_*)  ,
\label{R3}
\ee
 reaches the maximum  value, \be R_{max} \sim  \frac{1} {  e c n_T} \ln(T/\varepsilon_*), \label{Rm1}\ee at $n_m \sim n_T$ ($\mu \sim T$), and  decays inversely proportional to the concentration at  $n\to \infty .$  \cite{comment}

At high field or low temperatures   ($\varepsilon_* \gg T$), the dependence of $R$ on $n$ also has a maximum and is linear at small concentration.  However, as seen from Eq.~\eqref{R2}  there are  some  differences compared to the low-field case. First of all,   the coefficient  in the linear dependence at small $n$ turns out to be different
\be R \sim  \frac{ n }{ec n^2_T} \left(\frac{\varepsilon_*}{T}\right)^2.
\label{R4}
\ee
Secondly, the maximum value of $R$ is reached at much smaller concentration $n_m \sim (T/\varepsilon_*) n_T $  corresponding to $\mu \sim T^2/\varepsilon_* \ll T.$ Equation \eqref{R4} holds    below this concentration while at larger $n$ the Hall coefficient is given by conventional expression \eqref{R}. The maximum Hall coefficient reads
\be R_{max} \sim  \frac{1} {  e c n_T} \left(\frac{\varepsilon_*}{T}\right).\label{Rm2}\ee

\section{Charged impurities}
\label{ChargedImpurities}

In the previous sections we discussed magnetotransport in graphene with the short-range disorder.  One can see from the above derivations, that  the most interesting result, the  square-root  MR  at  low $H $  is a direct consequence of  the energy dependence of scattering time specific for the short-range impurities: $1/\tau_{tr} \propto |\varepsilon|.$
In this Section we discuss  scattering by the charged impurities  which are often considered  to give a dominant contribution to the resistivity
of graphene. In particular, the Coulomb impurities yield a linear dependence of the conductivity on the carrier concentration away from the Dirac point,
in agreement with the experimental data on most graphene samples.

The matrix element of the scattering on a single charged impurity is given by
\be
V_q=\frac{2 \pi e^2/q\varkappa}{1+(2 \pi e^2/q\varkappa)N\Pi(q)},
\label{polar}
\ee
where $\Pi(q)$ is the static  polarization operator and $\varkappa$ is the dielectric constant.
If we neglect the screening of the impurities [which corresponds to $\Pi=0$ in Eq.~\eqref{polar}], and use golden rule  for calculation of the transport scattering rate, we get $1/\tau_{tr} \propto |\varepsilon|^{-1},$ which implies that $\omega_c\tau_{tr}$ does not depend on energy and, consequently, the  mechanism of the MR discussed above does not work.  Note that,
  in contrast to conventional semiconductors, where condition $\omega_c\tau_{tr}=const$ guaranties absence of MR,  in graphene there should be a parabolic MR  (more pronounced for $\mu \le T$) within the  model neglecting  screening. This is because of a partial cancelation of the electron and hole contributions to the transverse
  conductivity. In particular, exactly at the Dirac point $\sigma_{xy}=0$ and hence $\rho_{xx}(H)=1/\sigma_{xx}(H)$, so that MR is given by ${ \Delta\varrho_{xx}}/{\varrho_{xx}(0)}= (\omega_c\tau_{tr})^2 \propto H^2.$

Let us now take screening into account.  We restrict ourselves to the analysis of the MR at the Dirac point, where one expects the most pronounced effect.
The general expression for polarization operator at the Dirac point  was derived in Ref.~\onlinecite{Schuett-ee}. For our purposes it is sufficient to know the asymptotical expression for $\Pi(q)$ at $q \ll T/\hbar v:$
\be
\label{polar1}
\Pi(q) \approx \frac{T \ln2}{\pi \hbar^2 v^2}, ~\text{for} ~q \ll T/\hbar v.
\ee
Indeed, as we discussed in the previous sections the main contribution to resistivity at weak fields  comes from the low energies, $\varepsilon \ll T.$ Since transferred  momentum $\hbar q$ is on the order of $\varepsilon / v ,$ we conclude that relevant momenta are smaller than $T/\hbar v.$

Further consideration depends on the value of the parameter $\alpha= e^2/\hbar v \varkappa$ which characterizes the strength of the electron-electron interaction. Whereas for free-standing graphene $\alpha \approx 2,$ it can be much smaller for graphene grown or placed on a substrate with large dielectric constant, as well as for graphene suspended in a media with large $\varkappa$ (for example, in conventional water). Moreover, $\alpha$ is suppressed due to the  renormalization of the Fermi velocity $v$. \cite{gonzalez11}
 Below we use $\alpha$ as a parameter of the theory and discuss separately two cases: $\alpha \sim 1$  and $\alpha \ll 1.$

 Before going to the calculations, we note that the screening of the Coulomb impurities is not the only effect of the electron-electron interaction on transport properties of graphene. It was
 shown in Ref.~\onlinecite{theory_Boltzmann} that inelastic collisions of carriers also produce a parabolic MR near the charge neutrality point.
 In what follows, we will first disregard inelastic collisions, taking into account only the screening effects (the role of inelastic
 collisions  will be briefly discussed  in Sec.~\ref{collisions}).
 We will see that the screening of Coulomb impurities changes the situation and, remarkably, in the absence of the inelastic collisions,
 the low-field MR becomes proportional to $\sqrt{H},$ in a full analogy with the  short-range scattering, though the temperature dependence of the MR is different.
 In what follows, we focus on the contribution of the overlapping Landau levels to the MR and analyze the conductivity semiclassically within the Drude-Boltzmann approach. As we mentioned above, such  approach allows one to obtain correct equation for MR up to a numerical coefficient.
 For simplicity, we will consider  the low-field asymptotic only.

\subsection{Strong electron-electron interaction: $\alpha\sim 1$ }
In this case, for $q \ll T/\hbar v$  we find
\be
V_q \approx \frac{1}{\Pi} \approx \frac{\pi \hbar^2 v^2}{N T \ln 2}.
\label{V}
\ee
We see that at low electron energies scattering matrix element does not depend on the energy just as in the case  of the short-range potential. Hence, in order to find low-field MR one should  make the following replacement $u_0 \to u_0'={\pi \hbar^2 v^2}/{N T\ln 2}$  and, consequently,
\be
\gamma \to  \gamma'=\frac{2N^2 (\ln2)^2 T^2}{n_\text{imp} \pi^2 \hbar^2 v^2}.
\label{ga}
\ee
Since $\gamma$ becomes temperature-dependent we conclude that energy $\varepsilon_*$ now also depends on $T$:
\be \varepsilon_* =\hbar\Omega \sqrt{\gamma'} \propto T \sqrt{H}.  \label{var*1}\ee
Low-field MR at the Dirac point is  still given by Eq.~\eqref{low-field},
where one should substitute  expression \eqref{var*1} for $\varepsilon_*.$
It is worth emphasizing  that
charged impurities yield temperature independent MR in contrasts to inverse temperature dependence of MR in the case of short range potential.

\subsection{Weak electron-electron interaction: $\alpha\ll 1.$ }
In this case, a new energy scale, $\alpha T $, appears within the temperature window.  For small energies, $ \varepsilon \ll \alpha T,$  the characteristic momentum transferred in the scattering event is on the order of $\varepsilon/\hbar v$ and
we get from Eq.~\eqref{polar}:   $V_q \approx 1/\Pi.$ Hence,  the potential is effectively short-ranged and its strength is
characterized by $\gamma^\prime$ [see Eq.~\eqref{ga}] in a full analogy with the previous subsection.

Next, we consider intermediate energies, $\alpha T \ll \varepsilon \ll T.$ Analyzing Eq.~\eqref{polar} one might conclude that the screening  can be neglected  and  $V_q$ is given by its value for unscreened Coulomb potential  $2\pi e^2/\kappa q.$  The issue, however, is more subtle  than it appears. Indeed,
estimates show that the bare Coulomb potential leads to an infrared divergency of the quantum scattering rate: $1/\tau_q \sim \int_0^k dq/q^2.$ This divergency is cured by screening, so that for a given electron energy $\varepsilon,$   the characteristic value of   $q$  turns out to be  much smaller than $\varepsilon/\hbar v,$ being on the order of $\alpha T/\hbar v$ (see also discussion in Ref.~\onlinecite{theory_Boltzmann}).  This, in turn, means that  the quantum scattering rate saturates when $\varepsilon$  becomes larger than $\alpha  T.$  On the other hand, for $\varepsilon\gg \alpha T$ the screening can be neglected  in the calculation of the transport scattering rate which is then determined by the bare Coulomb potential.  As a consequence,  $1/\tau_{tr}$ starts to decrease as $|\varepsilon|^{-1}$ for $\varepsilon > \alpha T$.

On a more formal level, expressing the transferred momentum $q=2(\varepsilon/\hbar v)|\sin(\theta/2)|$ through the scattering angle $\theta$,
we represent the scattering rates as
\begin{eqnarray}
     \frac{1}{\tau_{i}}
  &\sim& \alpha^2 v^2 \hbar n_\text{imp} \varepsilon  \int_0^\pi \frac{d \theta}{[\varepsilon \sin(\theta/2)+ \alpha N T \ln 2]^2} \nonumber\\
 &\times&
 \left\{
   \begin{array}{cc}
     1+\cos\theta, &\text{for} \quad i=q \\
     \sin^2\theta, &\text{for}\quad i=tr \\
   \end{array}
 \right. .
\end{eqnarray}
Evaluation of the integrals yields:
 \be
 \frac{\hbar}{\tau_{q}}\sim
 \left\{  \begin{array}{c}
                               \varepsilon/ \gamma^\prime, ~\text{for}~\varepsilon \ll \alpha T; \\
                               \alpha T/\gamma^\prime, ~\text{for}~\varepsilon \gg \alpha T.
                             \end{array}\right.
  \label{tauq1}
  \ee
 \be \frac{\hbar}{\tau_{tr}}\sim \left\{  \begin{array}{c}
                               \varepsilon/2\gamma^\prime, ~\text{for}~\varepsilon \ll \alpha T; \\
                               (\alpha T)^2/\gamma^\prime\varepsilon, ~\text{for}~\varepsilon \gg \alpha T.
                             \end{array}\right.
  \label{tautr1}\ee
 Dependence  of the transport and quantum  scattering rates on the energy  at small $\alpha$ is plotted schematically  in Fig.~7. The value of $\gamma^\prime$ is given by Eq.~\eqref{ga}.
 \begin{figure}[ht!]
 \leavevmode \epsfxsize=8.0cm
 \centering{\epsfbox{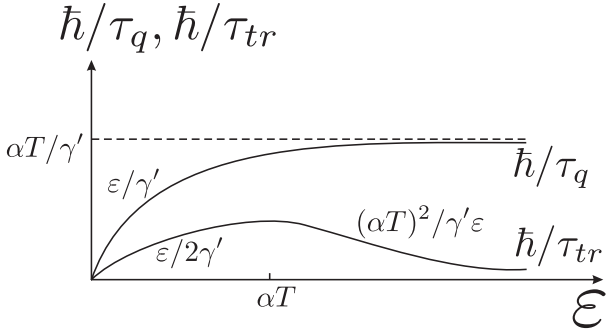}}
 \caption{Energy dependence of the quantum and transport scattering rates
 }
 \end{figure}

Let us now  consider conductivity at the Dirac point ($\mu=0$) within the Drude-Boltzmann approach.   Since most interesting results are expected at low fields, we restrict ourselves to discussion of the case $\varepsilon_* \ll \alpha T.$ Using Eqs.~\eqref{omega-c}, \eqref{eps*} and \eqref{tautr1} we find
\be \sigma_{xx}(\varepsilon)\sim \frac{e^2\gamma^\prime}{\hbar}\left\{  \begin{array}{c} \displaystyle
                               \frac{\varepsilon^4}{\varepsilon^4+\varepsilon_*^4}, ~\text{for}~\varepsilon \ll \alpha T; \vspace{1mm}\\ \displaystyle
                               \frac{\varepsilon^2 (\alpha T)^2}{(\alpha T)^4+\varepsilon_*^4}, ~\text{for}~\varepsilon \gg \alpha T.
                             \end{array}\right.
  \label{sigmaxx2}\ee
 Next we perform the thermal averaging  of $\sigma_{xx}(\varepsilon)$ and take into account that the averaged transverse conductivity at the Dirac point is zero due to the cancelation of the electron and holes contributions.
 Straightforward calculations yield the following result for MR:
\be
  \frac{ \Delta\varrho_{xx}}{\varrho_{xx}(0)} \sim \left\{  \begin{array}{c} \alpha^2 \varepsilon_*/T
                              \propto \sqrt{H}, ~\text{for}~\varepsilon_* \ll \alpha^2  T ; \\
                               (\varepsilon_*/\alpha T)^4\propto H^2, ~\text{for}~\varepsilon_* \gg \alpha^2  T.
                             \end{array}\right.
  \label{MR2}\ee
As seen, with increasing the magnetic field, the square-root dependence on magnetic field changes to a parabolic one.
It is notable  that due to the quadratic dependence of $\gamma^\prime$ on $T$ [see Eq.~\eqref{ga}], the MR turns out to be temperature independent both at weak  and at strong fields.

When deriving of equation \eqref{MR2}, it was assumed  that $\epsilon^*\gg\Omega,$ which implies that $\gamma^\prime \gg 1$ [see Eq.~\eqref{eps*}].
Actually, the latter condition is not crucial for our semiclassical treatment.
Indeed, at zero field, the energy-averaged longitudinal conductivity is given by $e^2\gamma^\prime/\hbar \alpha^2.$
This conductivity should be large compared to $e^2/\hbar,$ which yields   $ \gamma^\prime \gg \alpha^2.$  Another condition used
in the semiclassical analysis is $T\tau_{q}(T) \gg \hbar.$ This condition, which  ensures that the  density of states
is not changed essentially at typical energies,  leads to a stronger inequality [see  Eq.~\eqref{tauq1}]: $\gamma^\prime \gg \alpha.$
Hence, in contrast to the case of short-range disorder,  $\gamma^\prime$ can be smaller than unity. For $\alpha \ll \gamma^\prime \ll 1,$ the MR is parabolic and is given by the bottom line of Eq.~\eqref{MR2} in the whole range of magnetic fields addressed in the paper.

Finally we note that the potential of the charged impurities is also screened by the gate electrode. Such screening becomes important at low temperatures, when characteristic values of the transferred momentum, $q \sim T/\hbar v, $ become smaller than the inverse spacer distance, $1/d.$ In this case, the amplitude of the screened impurity potential is estimated as $V_q=2 \pi e^2d/\varkappa=const,$ so that this case is equivalent to the case of the short-range disorder.

\section{Role of inelastic collisions} \label{collisions}

So far, we have completely disregarded the effect of inelastic collisions induced by Coulomb interaction between the carriers.
In fact, such collisions  are crucially important for transport properties of clean graphene in the vicinity of the Dirac point ($\mu \ll T$).\cite{theory_Boltzmann,Schuett-ee,drag,Kashuba,Fritz, aleiner, fluid, Foster}
A detailed analysis of the MR in the presence of such collisions is out of the scope of the current research.
Here we limit ourselves to a qualitative discussion of the problem.

Let us first return to the case of short-range impurity potential.
For electrons with typical energies, $\varepsilon \sim T,$ the rate of inelastic collisions is of the order of $1/\tau_{ee}\sim\alpha^2 T/\hbar.$
This rate should be compared with the characteristic scattering rate off impurities [see Eq.~\eqref{tau0}], which yields for typical energies
$\epsilon\sim T$:  $1/\tau_{q}\sim 1/\tau_{tr}\sim T/\gamma \hbar.$
Since  $\gamma $ is large by assumption, the existence of the disorder-dominated regime requires rather weak interaction:
\be \alpha^2\ll 1/\gamma \ll 1 . \label{app}\ee
  In the opposite limit, $\alpha^2\gg 1/\gamma,$ the hydrodynamic approach developed in Ref.~\onlinecite{theory_Boltzmann}
  yields a parabolic MR.  As we already mentioned above, the interaction constant
 $\alpha= e^2/\hbar v \varkappa$ can be much smaller  than unity
 for graphene  grown on a substrate with large dielectric constant or for graphene suspended in a
 media with large $\varkappa$. Further, $\alpha$ is suppressed due to the renormalization of the Fermi velocity $v$. \cite{gonzalez11}
This implies that inequality $\alpha^2\ll 1/\gamma $ can be satisfied in experiments.

We now briefly discuss the competition between  the scattering off the charged impurities and the electron-electron collisions.
As follows from  Eq.~\eqref{tautr1}, for typical energies, $\varepsilon \sim T,$  scattering rate is estimated as $1/\tau_{tr} \sim \alpha^2 T/\hbar \gamma^\prime.$ Comparing this rate with the inelastic one, $1/\tau_{ee},$ we arrive at the conclusion that at $\gamma^\prime \gg 1$ the electron-electron collisions
 dominate over the impurity scattering. In this case, the MR  is  parabolic and described  by the theory developed in   Ref.~\onlinecite{theory_Boltzmann}.
In the notation used above, one can represent MR obtained in   Ref.~\onlinecite{theory_Boltzmann} for the case $\mu=0$ as follows:
\be
\frac{ \Delta\varrho_{xx}}{\varrho_{xx}(0)} \sim \omega_c(T)^2 \tau_{ee}(T) \tau_{tr}(T) \sim \left(\frac{\varepsilon_*}{\alpha T}\right)^4 \frac{1}{\gamma^\prime}. \label{muller}
\ee
We see that in the collision-dominated regime, inelastic collisions suppress the MR by a factor of $\gamma^\prime\gg 1$ as compared to
the collisionless case [see the bottom line in Eq.~\eqref{MR2}].

In the opposite limit, $\gamma^\prime \ll 1,$ the scattering by impurities dominates over the
 electron-electron collisions. Nevertheless, the low-field  asymptotic of MR [the top line of Eq.~\eqref{MR2}] is not realized. Indeed, the condition
 $\epsilon \tau_q/\hbar \gg 1, $ which ensures that the spectrum is not essentially affected by scattering, is equivalent  to the condition $\varepsilon \gg \alpha T/\gamma^\prime$ and, therefore,  is not satisfied in the region of relevant energies,
 $\varepsilon \sim \varepsilon_*\ll \alpha^2 T.$  However,  at high fields, the  MR is dominated by impurity scattering and is given by the bottom line of Eq.~\eqref{MR2}  provided that  $\gamma^\prime \gg \alpha$ (as we mentioned above, the latter condition ensures that the spectrum is not changed at typical energies $\varepsilon \sim T$).

Let us finally emphasize that the square-root low-field MR  due to the scattering off Coulomb impurities
can be obtained in the gated graphene. The gate  screens  the Coulomb potential, thus weakening both the impurity and  the electron-electron scattering.
Here we focus on the case of  low temperatures  $T\ll\hbar v/d, $  when for typical energies ($\varepsilon \sim T$) and
wavevectors ($q\sim T/\hbar v$) we get $V_q \sim e^2d/\varkappa$ and, consequently, find:
\BEA
 &&\frac{1}{\tau_q}\sim \frac{1}{\tau_{tr}}\sim\frac{T}{\hbar \gamma^{\prime\prime}},\\
&&\gamma^{\prime\prime} \sim \frac{\hbar^2 v^2}{n_{\rm{imp}}V_q^2} \sim\frac{1}{\alpha^2n_{\rm{imp}}d^2 },\\
&& \frac{1}{\tau_{ee}}\sim \frac{\alpha^2T}{\hbar} \left(\frac{Td}{\hbar v}\right)^2.
\EEA
Condition $\gamma^{\prime\prime} \gg 1 $ yields
\be
n_{\rm{imp}} \ll \frac{1}{\alpha^2 d^2}.
\label{cond-smalln}
\ee
Impurity scattering dominates over electron-electron collisions when $1/\tau_{tr} \gg 1/\tau_{ee},$ thus giving another restriction
\be
T\ll \hbar v \sqrt{n_{\rm{imp}}}.
\label{condTn}
\ee
It is easy to see that both inequalities, Eq.~(\ref{cond-smalln}) and (\ref{condTn}), can be satisfied in the limit of low temperatures
for sufficiently low concentration of impurities.
At such concentrations and temperatures, the MR is given by expressions derived  above for the short-range disorder with the replacement of
$\gamma$ by $\gamma^{\prime\prime}.$

It is worth noticing that the assumption about the absence of Shubnikov--de Haas oscillations limits possible temperatures from below:
$T\gg \hbar \Omega/\sqrt{\gamma^{\prime\prime}}$.  In combination with Eq.~(\ref{condTn}), this condition yields a parametrically large
range of temperatures for the square-root MR, provided that $\alpha d/l_H \ll 1$.
Thus we conclude, that the low-field square-root MR is a generic feature in a realistic gated setup in a wide range of experimentally accessible
parameters.

Above we presented estimates both for short-range disorder and  for charged impurities assuming that the electron energies are on the order of $T.$   In fact, the situation is more complicated because the main contribution to the square-root dependence obtained above comes from  small (untypical) energies ($\varepsilon \ll T$).  The analysis of electron-electron collisions for such  energies  is nontrivial and should take into account the plasmon-assisted scattering mechanism \cite{Schuett-ee} (though the plasmons are not important\cite{Kashuba,Fritz,Schuett-ee} for those transport properties that are
determined by energies on the  order of $T$).
The MR in the presence of interaction might still be determined by low untypical energies. Detailed study of the plasmon-assisted scattering and its competition with the impurity scattering
is a challenging problem which  will be discussed elsewhere.

\section{Discussion and conclusions} \label{con}
To conclude, we have studied   magnetotransport in graphene, focusing on the case of short-range disorder.
We have found  that MR depends on three relevant parameters having dimensionality of the energy, $\mu,T,$ and $\varepsilon_*,$
the field dependence being fully absorbed by $\varepsilon_*$ which is proportional to  $ \sqrt{H}.$
One of the   main predictions  of our model is the  square-root field dependence of MR in the limit of low $H$,
both at  very low and at very high temperatures. Such a  dependence persists down to
exponentially small fields, corresponding to    $\varepsilon_* \sim  \Delta e^{-\pi\gamma/2} .$

We separately analyzed the cases of low and high temperatures  and identified four  different transport
regimes for $\mu \gg T$  and three regimes for $\mu \ll T$ (see Fig.~4).
All these regimes can be realized provided that temperature lays within a certain interval. Let us now find the corresponding criteria.

For $\mu \gg T$ and not too small $\varepsilon_*,$ MR is determined by energies close to the Fermi
surface, $\varepsilon \approx \mu.$ Above we assumed  that  Shubnikov-de Haas oscillations are suppressed by energy
averaging within the temperature window, which implies that $T \gg \hbar \omega_c(\mu). $ Using Eqs.~\eqref{omega-c} and \eqref{eps*},
the latter inequality can be rewritten as $\varepsilon_* \ll \sqrt{\gamma T\mu}.$  While identifying regimes shown at the upper
picture in the Fig.~4 we implicitly assumed that $\sqrt{\gamma T\mu}$ is larger than $\mu,$ or,
equivalently, $T> \mu/\gamma.$ When $T$ becomes smaller than  $ \mu/\gamma$ the high-field square-root
asymptotic of MR is not realized  because of arising of
Shubnikov de Haas oscillations at $\varepsilon_* \sim \sqrt{\gamma T\mu} \ll \mu.$ However,
the  low-field square-root  asymptotic is determined by energies $\varepsilon \sim \varepsilon_*$
and remains valid even at low temperature, because Eq.~\eqref{ineqT}  is always satisfied for $T\ll \mu$ and $\varepsilon_* < \varepsilon_2.$

In the opposite limiting case, $T \gg \mu,$ the main contribution to  MR comes from $\varepsilon \approx \varepsilon_*.$
Rewriting Eq.~\eqref{ineqT}  as $\varepsilon_* \ll \gamma T,$ we find that three regimes shown in Fig.~4 are realized
provided that $\gamma T \gg T^2/\mu,$ or, equivalently, $T \ll\gamma \mu.$  At higher temperatures, $T\gg \gamma \mu,$ MR is
given by $C_3\varepsilon_*/2T$ for $\varepsilon_* \ll T$ and $C_4(\varepsilon_*/T)^3$ in the interval $T \ll \varepsilon_* \ll \gamma T.$
The Shubnikov-de Haas oscillations appear  at $\varepsilon_* \gg \gamma T.$

We  predicted a non-trivial behavior of the Hall coefficient on $H$ in the vicinity of the Dirac point.
With increasing the field, $R$ decreases, reaches a minimum and then starts to grow again. Further, we analyzed
dependence of $R$ on  electron concentration  and found that this dependence is non-monotonic both for law and strong fields.

We also estimated the MR caused by scattering off the charged, partially-screened impurities and discussed the competition between disorder- and collision-dominated  mechanisms of MR.
 Importantly, the main prediction of our theory, the square-root dependence of MR in the limit of weak fields,
is also valid at sufficiently low temperatures in the case of charged impurity potential when the screening by an external gate is taken into account. Specifically, in this situation, the range of $T$, where the effect takes place, is parametrically large for $\alpha d/l_H\ll 1$
which is easily accessible in experiments.



Before concluding the paper, let us discuss some interesting problems to be addressed in future. First, we note  that  our theory is applicable to any other 2D electron
systems  with the linear  energy spectrum. Simplest examples are surfaces of 3D topological insulators   and  2D spin-orbit metals  based, in particular, on the CdTe/HgTe quantum wells with the chemical potential moved away from the gap.  \cite{chto_to_obscheje_pro_Hg,zhang}
The latter example needs some comments.   The energy spectrum in  2D spin-orbit metals is not purely linear  (with the only exception  for a certain width of  the quantum well), so that the evident generalization of our theory is the calculation of the MR for the case of a small but finite effective mass $m$  of the carriers. Besides 2D spin-orbit metals,  such a theory would apply to a bilayer graphene.   We expect  that for a quasi-linear  spectrum, MR would depend on the relation between $T$ and $mv^2.$ In the most interesting case, $\mu \ll T,$  we expect  parabolic MR at $T \ll m v^2$  which  would cross over to the square-root MR at $T\gg mv^2.$
Further, our work motivates the following question related to the interaction effects in graphene:
What is the effect of inelastic scattering (including the role of plasmons) on the Landau-level broadening in graphene?
In particular, in a clean sample, Coulomb interaction is the only source of such broadening
(note that it is this broadening that justifies a semiclassical hydrodynamic approach).
A square-root MR might still arise in the presence of inelastic scattering in graphene for sufficiently strong magnetic fields,
such that within the temperature window both regions of separated and overlapping Landau levels
are present, as in the disorder-dominated regime addressed in our work.
Finally, it is interesting to analyze the MR of a suspended graphene, where scattering by flexural phonons
is crucially important for the transport properties. \cite{flexural}

Note added: After this work was completed, the experimental evidence of square-root MR in monolayer graphene was reported.\cite{alekseev}
Further, very recently, the \emph{ac} magnetoconductivity was calculated in Ref.~\onlinecite{Briskot} for the case of point-like
impurities in graphene; the results of Ref.~\onlinecite{Briskot} in the limit $\omega\to 0$ are in agreement with our predictions.

\section{Acknowledgements}
We thank N.S. Averkiev, A.A. Greshnov, {B.N. Narozhny}, M.O. Nestoklon, {R. Haug}, M. Titov, {G. Yu. Vasil'eva},  and Yu. B. Vasilyev  for
fruitful discussions.
The work was  supported   by DFG CFN, BMBF, DFG SPP 1459 ``Graphene'', RFBR,
RF President Grant "Leading Scientific Schools" NSh-5442.2012.2,   by  programs of the RAS, and by the Dynasty Foundation.


\begin{thebibliography}{100}
\bibitem{electrons} The factor $\mu/T$  in the averaged value of $\sigma_{xy}^D$ appeared due to the partial cancellation of the electron and hole contributions which have  different signs.  This factor is absent in an artificial model where there is only one sort of Dirac carriers (say, electrons). It is worth noting that even in this model, the averaged transverse conductivity, $\sigma^D_{xy}\sim |\Delta\sigma_{xx}(H)|,$ satisfies Eq.~(\ref{ineq-sigmaxy}) and the estimate \eqref{est} holds, giving rise to a square-root MR.
\bibitem{koechto_0} Within the SCBA, the transverse conductivity including both ${\sigma}_{xy}^{\uppercase\expandafter {\romannumeral 1}} (\varepsilon)$ and ${\sigma}_{xy}^{\uppercase\expandafter {\romannumeral 2}} (\varepsilon)$ can be written as follows \cite{ando-xy} $$\sigma_{xy} (\varepsilon)=\frac{ecn(\varepsilon,H)}{H} -\frac{\sigma_{xx}(\varepsilon) z(\varepsilon)}{\omega_c(\varepsilon)\tau_{tr}(\varepsilon)}  .$$  This equation is obtained from the semiclassical   Drude-Boltzmann relation between $\sigma_{xx}(\varepsilon)$ and $\sigma_{xy}(\varepsilon)$ by replacement of $\tau_{tr}$  with $\tau_{tr}/z.$ Subtracting ${\sigma}_{xy}^{\uppercase\expandafter {\romannumeral 1}} (\varepsilon)$ [given by Eq.~\eqref{sigmaxy0}] from this relation, one finds a convenient representation for thermodynamical contribution to the Hall conductivity
    $${\sigma}_{xy}^{\uppercase\expandafter {\romannumeral 2}} (\varepsilon)=\frac{e^2N}{2\pi\hbar^3\Omega^2}\left[ \int_0^\varepsilon z(\varepsilon')|\varepsilon'|d\varepsilon'- \frac{|\varepsilon|\varepsilon z(\varepsilon)}{2}\right].$$ Averaging the latter equation over energy within the interval $\delta \varepsilon$ one finds that the thermodynamical contribution can be neglected.
\bibitem{koechto}
In the energy range $\varepsilon\gg \varepsilon_*$ corresponding to overlapping Landau levels,
$\sigma_{xy}^{\uppercase\expandafter {\romannumeral 2}} (\varepsilon)$ is exponentially small and besides that is an oscillatory
function of energy, whereas the exponentially small correction to  $\sigma_{xy}^{\uppercase\expandafter {\romannumeral 1}} (\varepsilon)$
contains a non-oscillatory term. Therefore,  energy-averaged  contribution to the transverse conductivity,
${\bar\sigma}_{xy}^{\uppercase\expandafter {\romannumeral 2}} (\varepsilon),$ is smaller than the exponential correction to ${\bar\sigma}_{xy}^{\uppercase\expandafter {\romannumeral 1}} (\varepsilon)$.
\bibitem{comment} One can show that crossover from Eq.~\eqref{R3} to Eq.~\eqref{R} happens in the interval of concentration corresponding to $T< \mu < T \ln[\ln (T/\varepsilon_*)].$
\end{thebibliography}
\end{document}